\documentclass[prb,aps,reprint,showpacs,amssymb, superscriptaddress,footinbib]{revtex4-1}
\usepackage{graphicx}
\usepackage{amsmath}
\usepackage[utf8]{inputenc}
\usepackage{amssymb}
\usepackage{color}
\usepackage{wasysym}
\definecolor{green0}{rgb}{0,0.5,0}

\begin{document}

\title{Emergence of Mixed Quintet Superfluidity in the Chain of Partially
  Polarized\\ Spin-3/2 Ultracold Atoms}

\author{G. Barcza }
\email[electronic address: ]{barcza.gergely@wigner.mta.hu}
\affiliation{MTA-WRCP Strongly Correlated Systems ``Lend\"ulet" 
Research Group, H-1525 Budapest, Hungary}
\affiliation{Department of Physics, Lor\'and E\"otv\"os University, H-1518 
Budapest, Hungary}

\author{E. Szirmai} \email[electronic address: ]{eszirmai@gmail.com}
\affiliation{Wigner Research Centre for Physics, Hungarian Academy of
  Sciences, H-1525 Budapest, Hungary} \affiliation{BME-MTA Exotic
  Quantum Phases ``Lend\"ulet" Research Group, H-1111 Budapest,
  Hungary }

\author{\"O. Legeza} 
\affiliation{MTA-WRCP Strongly Correlated Systems ``Lend\"ulet" Research Group, 
H-1525 Budapest, Hungary}
\affiliation{Fachbereich Physik, Philipps Universit\"at, G-35032 Marburg, Germany}

\author{J. S\'olyom} 
\affiliation{Wigner Research Centre for Physics,
  Hungarian Academy of Sciences, H-1525 Budapest, Hungary}

\date\today

\begin{abstract}
The system of ultracold atoms with hyperfine spin $F=3/2$ might be
unstable against the formation of quintet pairs if the interaction is
attractive in the quintet channel.  We have investigated the behavior
of correlation functions in a model including only $s$-wave
interactions at quarter filling by large-scale density-matrix
renormalization-group simulations. We show that the correlations of quintet 
pairs become quasi-long-ranged, when the system is partially polarized, 
leading to the emergence of various mixed superfluid phases in which BCS-like 
pairs carrying different magnetic moment coexist. 
\end{abstract}


\maketitle


{\bf Introduction:} Recently ultracold atomic and molecular systems have been in the focus
of theoretical and experimental studies not only in atomic and molecular 
physics but also in condensed matter physics.\cite{lewenstein07a}
Atomic systems with hyperfine spin degrees of freedom higher than $1/2$ can
show completely new behavior for both bosonic and fermionic systems.
For repulsive interaction Mott insulating phases, (chiral)
spin liquid states, resonating plaquette order, spin-quadrupole and
even higher multipole order~\cite{wu2003a,harada03a,gorshkov10a}  or  a generalized
Peierls-like distortion can occur.\cite{buchta07a} For attractive
interactions bound trionic and quartet states have been predicted for
one- and three-dimensional systems as well.\cite{wu05a,rapp07a}

In addition to the usual singlet BCS pairs, non\-singlet pairs may
also occur if the appropriate component of the
interaction is attractive. A general description of different high-spin pair states 
was given 
by Ho and Yip.\cite{ho99a} $F=3/2$ 
fermions may form spin-2 (quintet) pairs, 
which are of
particular interest owing to their exotic properties.\cite{wu10}
Moreover ultracold atomic systems of $F=3/2$ fermions are excellent candidates for studying 
the consequences of high symmetries, since they possess  
SO(5)
\cite{wu2003a,wu2003b}

For a long time magnetic ordering and superconductivity were thought
to be incompatible. In fact homogeneous ferromagnetic order excludes 
homogeneous 
singlet superconductivity. 
Coexistence is possible for $p$-wave triplet pairs \cite{ferro} in 
crystallographically layered systems \cite{pnictides}
or in inhomogeneous singlet superconductivity with finite momentum of
the pairs, that is in the Fulde-Ferrell-Larkin-Ovchinnikov (FFLO)
phase.\cite{fulde64a}  This 
phase has been intensively
studied in two-component, spin polarized systems \cite{fflo, batrouni07}
and recently it has been realized in experiments
with a one-dimensional array of ultracold atoms by the Hulet group.\cite{ovachkin}

In the present paper, we study the possible formation of local quintet pairs
and their stability in a one-dimensional chain of fermionic atoms with
hyperfine spin $F=3/2$, when the interaction is attractive in the
quintet channel.  We consider a quarter-filled system, that is the
number of particles is equal to the number of sites. We show that
if spin states with different
spin components are unequally populated, 
quintet pairs can become stable.
Note that in a one-dimensional model, where no true long-range order may exist,
a superfluid state is claimed to be stabilized, when the corresponding
correlation function shows algebraic---instead of exponential---decay.
The superfluid phases will be characterized by the spin quantum number of 
the pairing operators appearing in the correlation function.
As will be seen, the type of stable quintet pairs depends on the 
coupling constants (scattering lengths) and the  spin-imbalance. 
It is worth noting that our model contains only $s$-wave interaction
indicating that quintet pairing phases can be stabilized via $s$-wave
Feshbach resonance.  This can help in the possible experimental 
realization of Cooper-like pairs with high 
multiplicity by avoiding the difficulties due to inelastic loss in
$p$-wave scatterings.

{\bf Formulation of the problem:} The scattering processes between particles with
hyperfine spin $F$ can be classified into independent spin channels
characterized by the total spin ($S$) of the 
scattered atoms. Accordingly, the interaction 
part of the Hamiltonian is
$\tilde V=\sum_{S=0}^{2F} g_S P_S$, where $P_S$ projects onto the total spin
$S$ subspace and $g_S$ is the coupling constant in the corresponding
channel. The projectors are expressed via
the pairing operators as $P_S=\sum_{m,i} P_{Sm,i}^\dagger
P_{Sm,i}^{\phantom\dagger}$, which are defined through the
Clebsch-Gordan coefficients \cite{clebsch} and the creation operator of fermions, $c_{\alpha,i}^\dagger$, 
at site $i$, with spin component $\alpha$ ($\alpha=\pm3/2,~ \pm1/2$), as $P_{Sm,i}^\dagger = 
\sum_{\alpha,\beta} \left< \frac32, \frac32;
\alpha, \beta |S,m\right> c_{\alpha,i}^\dagger c_{\beta,i}^\dagger$,
where $m$ is the $z$ component of the total spin of the two scattering
particles.
Starting from a fermionic spin-3/2 Hubbard-like model with
on-site interaction, the only contributing terms are antisymmetric
under the exchange of the spin of the two colliding atoms, therefore,
only the $S=0$ and $S=2$ terms may appear. 
Thus the Hamiltonian of the
system reads as
\begin{equation}
 \label{eq:ham1}
{\cal H} = -t\sum_{i,\alpha} \big( c_{\alpha,i}^\dagger
c_{\alpha,i+1}^{\phantom\dagger} + \mathrm{H.c.} \big) + g_0 P_0 + g_2 P_2,
\end{equation}
where $t$ measures the hopping amplitude between neighboring sites. 
In optical lattices $t \approx 2 \omega_{\mathrm{R}}  \zeta^3 e^{-2\zeta^2}/\pi$, 
with $\zeta = (V_0/ \omega_{\mathrm{R}})^{1/4}$, where $V_0$ is the potential depth, 
and $\omega_{\mathrm{R}}=\hbar^2 k^2/2m_{\rm atom}$ is the recoil energy 
for atoms of mass $m_{\rm atom}$
in a lattice with lattice constant $a$. The 
coupling $g_S$ 
measured in units of the transverse confinement energy 
is related to $V_0$ and the $s$-wave scattering length in the total spin 
$S$ scattering channel, $a_S$, as $g_S\approx \pi^2 \zeta a_S/(2a)$.
For attractive couplings ($g_0<0,g_2<0$) the above Hamiltonian suggests
that singlet and quintet pairs are competing. Although the SU(4)
line ($g_0 = g_2$) is expected to be the most relevant experimentally,
the region with repulsive interaction in the singlet channel ($g_0>0$) 
and attractive interaction in the $S=2$ channel turns out to
be more  favorable for quintet pairing. Therefore, besides the SU(4) line, we will consider 
the case $g_0>0,g_2<0$, where quintet pairing competes with density waves,
as seen if the interaction term \eqref{eq:ham1} is rewritten in terms of
the density, $n_i=\sum_\alpha n_{\alpha,i}=\sum_\alpha
c_{\alpha,i}^\dagger c_{\alpha,i}^{\phantom\dagger}$, and the $P_2$
quintet projector as $ U/2 \sum_i n_i^2 + V P_2$,
with couplings $U=2g_0$, $V=g_2 - g_0$. 
In a system with more than two components, not only pairs, but trions, too, may be formed. 
We did not consider such a possibility.
Although three-body losses \cite{3body} may be important in this system,
large three-body losses may suppress threefold occupation of sites and may stabilize 
pairs as has been shown in Ref. \onlinecite{3body-kantian}. 
 
Analytic calculation in the weak-coupling
limit \cite{wu05a} shows that the leading instability for $g_0 > 0$,
$g_2<0$ is the formation of site- or bond-centered spin singlet quartets, which are formed 
from an equal number of atoms with $\alpha=\pm1/2$, $\pm3/2$.
In order to search for possible conditions that might stabilize the quintet
Cooper pairs in one dimension we have studied numerically the phase
diagram of model \eqref{eq:ham1} for $g_2 \leq 0$ at quarter filling.

{\bf Numerical procedure:} 
Density-matrix renormalization-group (DMRG) \cite{dmrg} 
simulations have been performed with open boundary
condition up to $L=64$ sites, keeping 500--2000 block states and
using up to 8 sweeps. Properties of various phases have been
determined by analyzing the spatial variation of correlation functions
of different pairs, $\chi_{Sm}(i) = \langle P_{Sm,1}^{\dagger}
P_{Sm,1+i}^{\phantom\dagger} \rangle$, with $m=0$ for $S=0$, and $m=0$, $\pm1$, $\pm2$ 
for $S=2$, of quartets, $\chi_{Q}(i) = \langle Q_{1}^{\dagger} Q_{1+i}^{\phantom\dagger}
\rangle$ with $Q_{i}^{\dagger}=c^{\dagger}_{3/2,i}c^{\dagger}_{1/2,i}c^{\dagger}_{-1/2,i}c^{\dagger}_{-3/2,i}$, 
as well as density and spin-density correlation functions, $\chi_{n}(i) = \langle n_{1} 
n_{1+i} \rangle - \langle n_{1} \rangle \langle n_{1+i} \rangle$, $\chi_{\tilde m}(i) = \langle \tilde
m_{1} \tilde m_{1+i} \rangle - \langle \tilde m_{1} \rangle \langle
\tilde m_{1+i} \rangle$, where $\tilde m_i = \sum_\alpha \alpha
n_{\alpha,i}$. In the rest of the paper---for better visibility---only the
quintet pairing correlation functions $\chi_{2m}$ are shown in the figures 
for the five $m$ values.

\begin{figure}[t]
    \centering
	\includegraphics[scale=0.45]{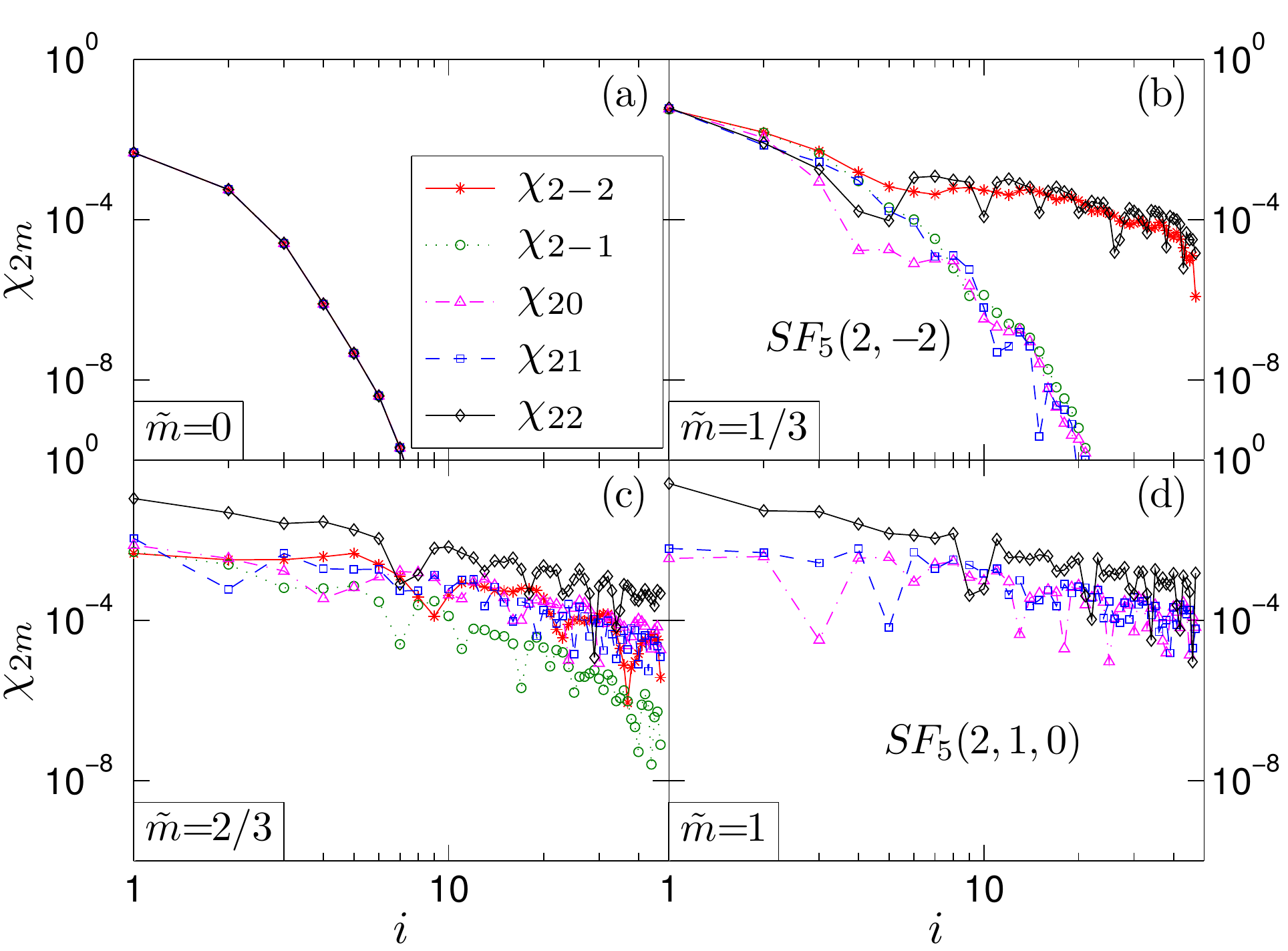}
    \caption{(Color online) The quintet correlation functions 
      $\chi_{2m}$ as a function of the distance $i$ for $\tilde{m}=0$ (all $\chi_{2m}$ 
      are equal), $1/3$, $2/3$  and $1$ calculated at $g_0=g_2=-1$.
      }
    \label{fig:P_vs_m_su4_line}
\end{figure}

\begin{figure}[t]
    \centering
	\includegraphics[scale=0.45]{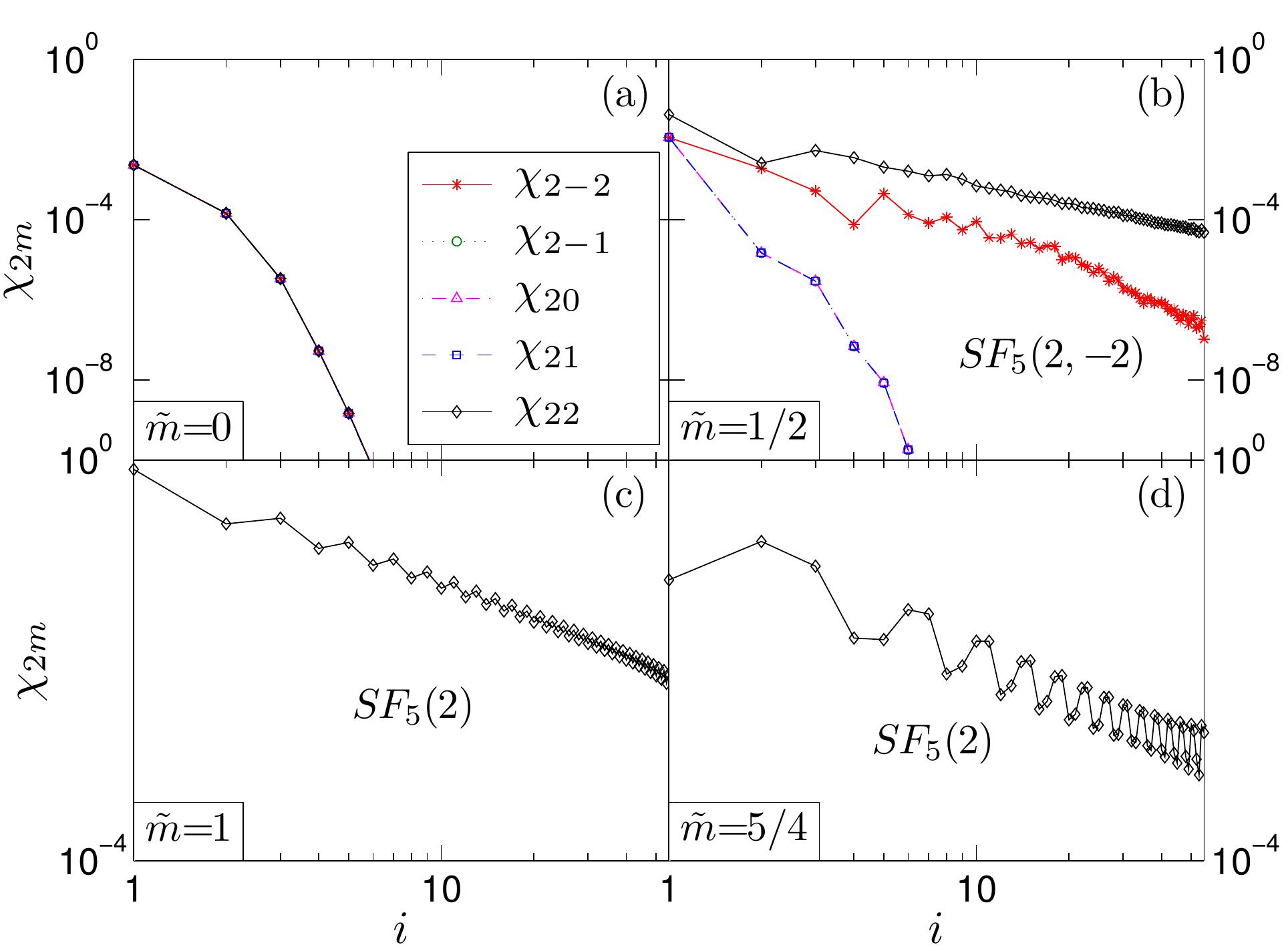}
    \caption{(Color online) 
      The same as Fig.~\ref{fig:P_vs_m_su4_line} but
      for $\tilde m=0$ (all $\chi_{2m}$ are equal), $\tilde m=1/2$ 
      ($\chi_{2,\pm2}$ are algebraic), $\tilde m=1$ (only $\chi_{22}$ is finite) and
      $\tilde m=5/4$ (only $\chi_{22}$ is finite) calculated for $g_0=2$, $g_2=-4$.
      }
    \label{fig:siteq}
\end{figure}

\begin{figure}[t]
    \centering
	\includegraphics[scale=0.45]{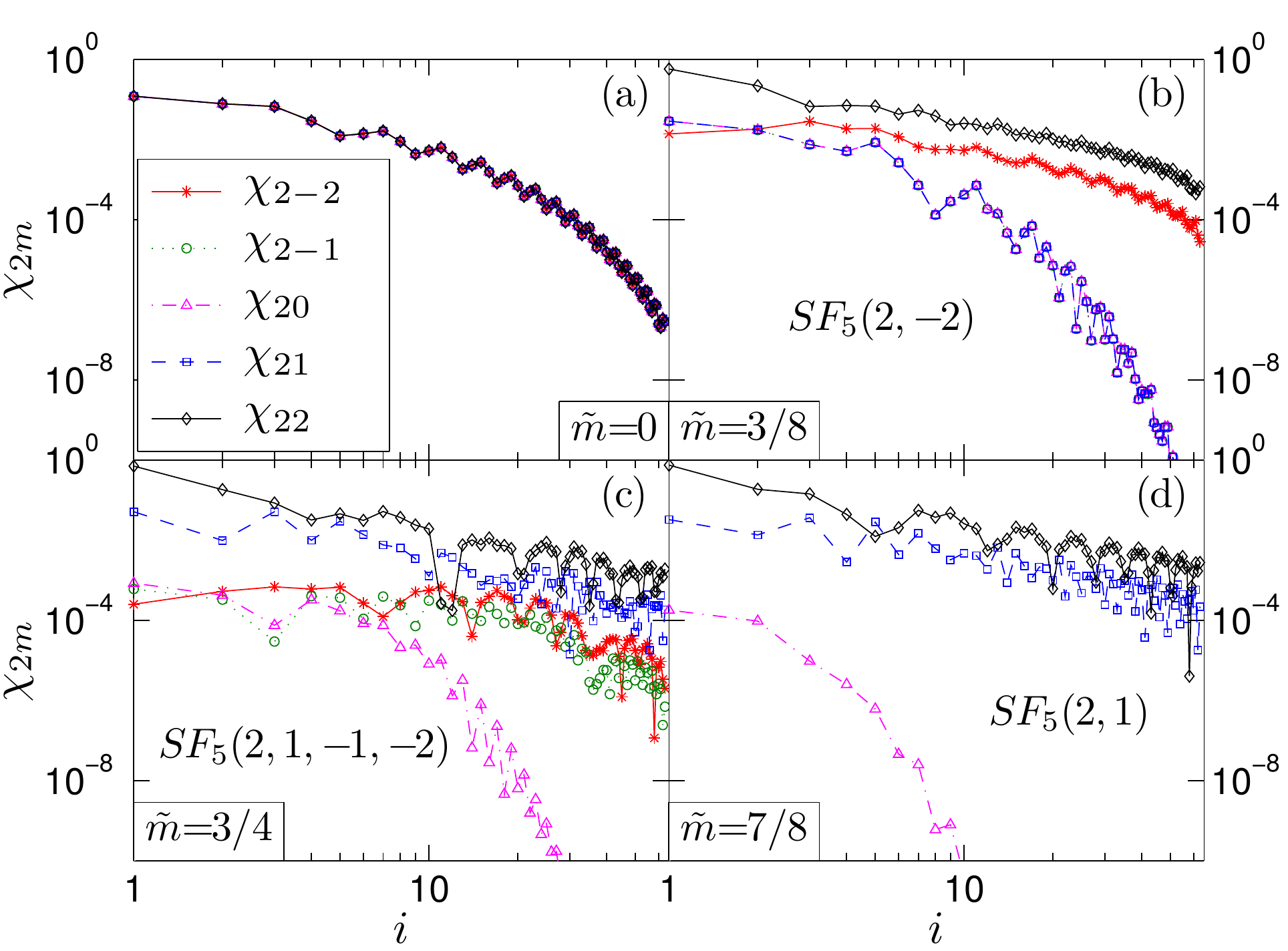}
    \caption{(Color online) 
      The same as Fig.~\ref{fig:P_vs_m_su4_line} but
      for $\tilde{m}=0,$ $3/8$, $3/4$ (when 
      $\langle n_{\alpha,i}\rangle\neq 0$ for all $\alpha$) 
      and $7/8$ (when $\langle n_{-3/2,i}\rangle$ vanishes) calculated 
      at $g_0=20$, $g_2=-4$. 
      }
    \label{fig:P_vs_m}
\end{figure}

{\bf Numerical results:} The analysis of these functions confirmed the
absence of quintet Cooper pairs when all spin components are equally populated,
 since $\chi_{2m}$ decays exponentially for all $m$ 
[see Figs. \ref{fig:P_vs_m_su4_line}(a), \ref{fig:siteq}(a), and \ref{fig:P_vs_m}(a)].  
We have found, in agreement with Ref.~\onlinecite{wu05a}, that
$\chi_{Q}$ and $\chi_{n}$ decay algebraically in the regime where a phase
composed of site-centered quartets was predicted by weak-coupling analysis.

The spin-singlet quartets could, however, be broken and quintet pairs could
be stabilized, if a population 
imbalance occurs in the number of fermions with different spin
components. Spin imbalance can for example be
generated by switching on a
weak magnetic field ($B$) which couples linearly to the magnetization
$\tilde m=\frac{1}{L}\sum_i \langle \tilde m_i \rangle$ (measured in
units of Bohr magneton). 
The stability of quintet pairs is indicated by the slow decay
of the correlation function of quintet pairs. Even for reasonably small spin imbalance,
 $\chi_Q$  decays   faster or vanishes, while 
the correlation of singlet Cooper pairs ($\chi_{0,0}$) behaves in the same way as $\chi_{2,0}$.
Therefore, in what follows we present the correlation
functions $\chi_{2m}$ for increasing spin imbalance from $\tilde m=0$ up to the
maximum value $\tilde m=3/2$.

{\it SU(4) symmetric model:} 
First we present results for the SU(4) symmetric model
where we have found three different superfluid phases as the polarization
is increased. Although it is difficult to determine the phase boundaries 
explicitly the correlation functions behave differently
for small ($\tilde m$ around 1/3), intermediate ($\tilde m$ somewhat below 1), and for large ($\tilde m>1$)
values of $\tilde m$, as can be inferred from Fig.~\ref{fig:P_vs_m_su4_line}. 
For $\tilde m=1/3$, the correlation functions for quintet pairs with $m=\pm 2$ 
show the slowest decay (Fig.~\ref{fig:P_vs_m_su4_line}(b)). We denote this phase by SF$_5(2,-2)$, 
where the subscript $5$ indicates the quintet nature of the superfluid (SF) phase and 
the numbers in brackets gives the $m$ index of the dominant $\chi_{2m}$ 
correlation functions. In this regime, the correlation function of the quartets decays only 
slightly faster than that of  
the dominant quintets. Note, that the occurrence of this state is highly nontrivial, 
since naively one can expect the dominance of pairs formed of fermions with  the majority spin
components $\alpha=3/2$ and $1/2$, while pairs formed of 
fermions with $\alpha=-1/2$ and $-3/2$ would be suppressed.

A different behavior is found at $\tilde m=1$  (Fig.~\ref{fig:P_vs_m_su4_line}(d)), where the 
correlation function of quintet pairs with $m=2$, 1, and 0 ($\chi_{2,2}$, 
$\chi_{2,1}$ and $\chi_{2,0}$) shows algebraic decay. 
The correlation functions $\chi_{2,-1}$ and $\chi_{2,-2}$, and also $\chi_Q$ vanish within our 
numerical accuracy, since the number of fermions with $\alpha = -1/2$ and $=-3/2$ is much less than 
that with $\alpha=3/2$. Similar reason is behind the smaller weight of the $\chi_{2,1}$ and $\chi_{2,0}$ 
quintet pairs, but they decay 
algebraically with the same exponent as $\chi_{2,2}$. The corresponding phase is 
denoted by SF$_5(2,1,0)$. 

The magnetization value $\tilde m=2/3$ (see Fig.\ref{fig:P_vs_m_su4_line}(c)) 
belongs to a region where the system possesses a transitional 
behavior between SF$_5(2,-2)$ and SF$_5(2,1,0)$. Here it is difficult to decide whether 
some of the correlation functions decay algebraically or exponentially. Nevertheless, it is 
clear that the dominant superfluid instability in this region is again characterized by 
different coexisting quintet pairs. 

Slightly above $\tilde m=1$, an effective two-component system with the usual FFLO state
develops, where only $\chi_{2,2}$ is finite and all other $\chi_{2m}$ are zero. 
This phase is denoted by SF$_5(2)$ and will be discussed in more detail below.

\begin{figure}[t]
    \centering
        \includegraphics[trim = 0mm 115mm 0mm 0mm, clip,scale=0.45]{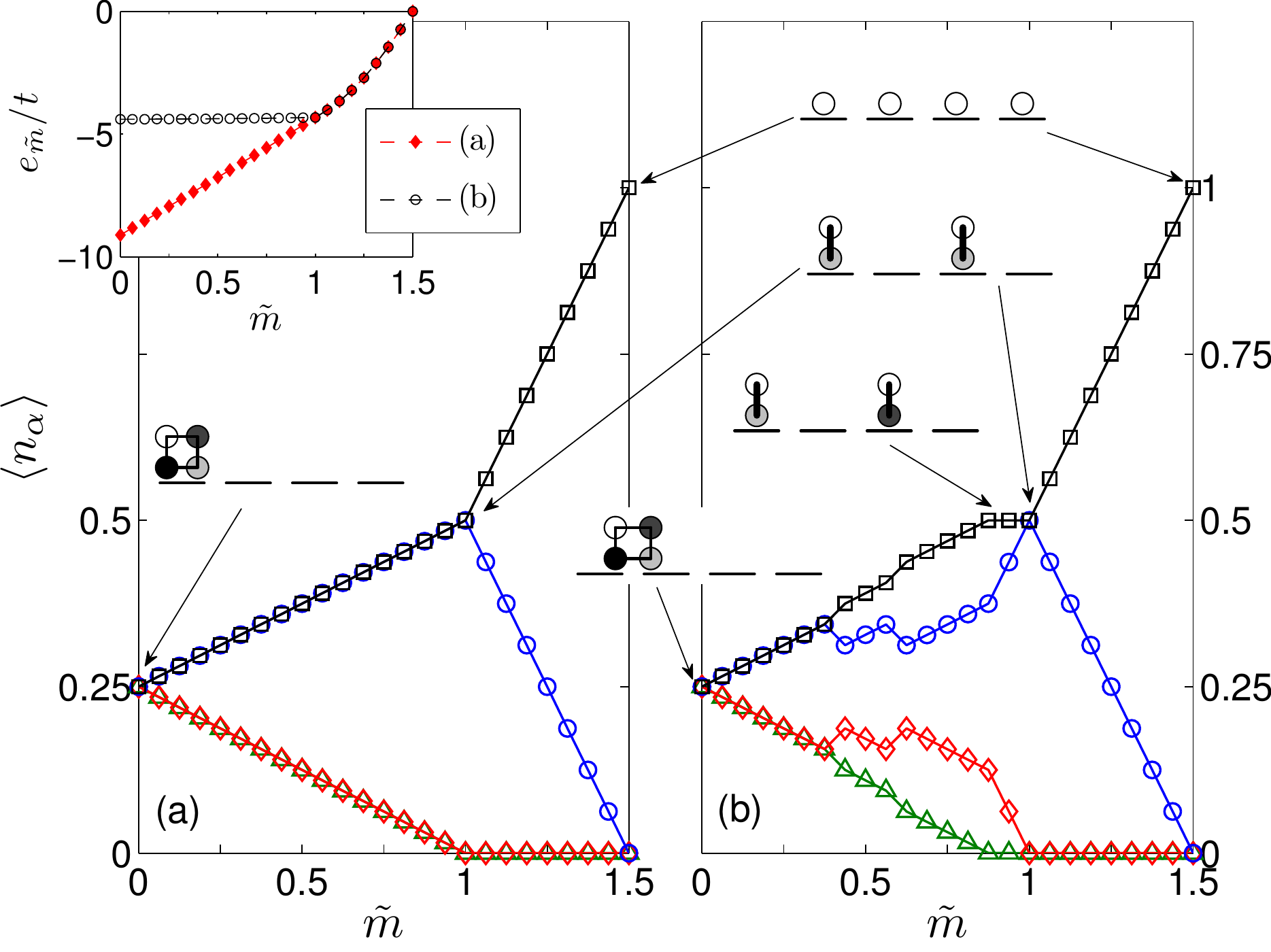}
    \caption{(Color online) Density of the fermions with different
      spin component vs. dimensionless magnetization $\tilde m$ for a chain with
      $L=64$ sites (a) at $g_0=2$, $g_2=-4$, and (b) at $g_0=20$,
      $g_2=-4$. The symbols {\color{green0}{$\vartriangle$}}, {\color{red}{$\Diamond$}}, 
     {\color{blue}{$\ocircle$}} and {\color{black}{$\Box$}}
      stand for $n_{-3/2}$, $n_{-1/2}$, $n_{1/2}$ and $n_{3/2}$, respectively.  
      The pictograms illustrate the  structure of
      the quarteting, quintet pairing and ferromagnetic phases (see
      the text for the details), where the shading of the circles indicates
      the hyperfine spin components of the atoms, $\alpha$,
      and the lines connecting the atoms indicate that the atoms form
      pairs or quartets with finite expectation values:
      $\big<c^{\dagger}_{\alpha}c^\dagger_{\beta}\big>$ and
     $\big<c^{\dagger}_{\alpha}c^\dagger_{\beta}c^{\dagger}_{\gamma}c^\dagger_{\delta}\big>$,
      respectively. The inset shows the
      site energy, $e_{\tilde m}$, as a function of $\tilde m$ for (a) and (b).}  
    \label{fig:sumN_L64}
\end{figure}

{\it The $g_0>0$, $g_2<0$ quadrant:} 
The calculation for the SU(4) symmetric model shows that various exotic 
mixed superfluid phases can exist in which BCS-like 
pairs carrying different magnetic moments coexist.
The differing behavior of the correlation functions is related to the number 
of atoms needed to form the pairs. Therefore, it is interesting to see how 
the number of atoms with $\alpha= -3/2, -1/2,$ and $1/2$
decreases while more and more atoms have $\alpha=3/2$ as the total polarization of the system is increases.
These numbers depend on the interaction between the particles.
For the sake of convenience, in what follows, we will consider the quadrant
$g_0>0$, $g_2<0$, because the decay of correlation functions is easiest to analyse there. 
We have found two types of dependence of
$\left<n_{\alpha,i}\right>$ on $\tilde m$ as displayed in
Fig.~\ref{fig:sumN_L64}. The regions, where one or the other behavior
is realized, correspond roughly to the regions separated by the line $g_0=-3g_2$, where 
the ground state at $\tilde m=0$ is a site-centered or  bond-centered phase.~\cite{wu05a}  
The difference is also apparent in the different behavior
of the site energy, $e_{\tilde m}$, as a function of $\tilde m$ 
(see the inset in Fig.~\ref{fig:sumN_L64} (a)).

In the whole quadrant $g_0>0,g_2<0$ the model becomes 
independent of $g_0$ for $\tilde m\geq 1$ due to the
absence of fermions with $\alpha=-3/2$ and $-1/2$.
The only surviving quintet correlation function, $\chi_{22}$, shows an algebraic decay
as shown in Fig.~\ref{fig:siteq}(c) and (d) in agreement
with the results of Batrouni $et~al.$ \cite{batrouni07}, since in the  $\tilde{m} \geq 1$ regime
our model can be mapped exactly to their two-component model. 
This phase, in our notation SF$_5(2)$, is equivalent to the 
well-known FFLO state. 

In contrast to this,  for $\tilde m<1$ the population imbalance of fermions with
different spin components shows markedly different character in the
two regions of the coupling space.  For $g_0<-3g_2$
(Fig.~\ref{fig:sumN_L64}(a)) all spin components 
have finite
weight for $\tilde m<1$.  As a consequence, the density of spin
quintet pairs 
decreases, but $\chi_{22}$ remains the slowest
decaying correlation function  at least when
$\tilde{m}\ge1/2$ (see Fig.~\ref{fig:siteq}(b)). In addition, also
$\chi_{2,-2}$ decays algebraically, although, with smaller weight 
and somewhat larger exponent. This mixed phase is the same SF$_5(2,-2)$ phase which was found 
along the SU(4) line for moderate magnetizations. Although, it is difficult to distinguish between an
exponential or algebraic decay of the correlation functions below
$\tilde m\approx 1/2$, quintet pairing is still the dominant
instability even slightly below $\tilde m=1/2$. 
 Even though all the four spin components have finite weight, the other correlation
functions $\chi_{2m}$ with $m=0$, and $\pm 1$
decay exponentially (Fig.~\ref{fig:siteq}(b)) for $\tilde m<1$. 

A different behavior is found for $g_0>-3g_2$
(Fig.~\ref{fig:sumN_L64}(b)). As the polarization decreases from $\tilde
m=1$ to a value slightly above 3/4, the density of atoms with $\alpha=-3/2$ remains
zero,
half of the atoms have $\alpha=3/2$, while the density of atoms
with $\alpha=\pm 1/2$ varies linearly with $\tilde m$.
$\chi_{2,-2}$ and $\chi_{2,-1}$ are equal to zero in this range 
and $\chi_{20}$ decays exponentially (Fig.~\ref{fig:P_vs_m}(d)).  On the other hand, the slowest
decaying correlation functions, $\chi_{21}$ and $\chi_{22}$, show
algebraic decay with identical exponent, therefore we call this phase SF$_5(2,1)$. 
As $\left<n_{-1/2}\right>$ is increasing, the
density of $m=1$ pairs also increases and the number of $m=2$
pairs decreases. At $\tilde m=3/4$ we have found that the $m=1$ and $m=2$ quintet pairings remain the
dominant instability, however, as $\tilde m$ is decreasing, the correlations $\chi_{2,-1}$ 
and $\chi_{2,-2}$ start to increase (see Fig.~\ref{fig:P_vs_m}(c)). 
For even weaker polarization again the SF$_5(2,-2)$ state is stabilized (Fig.~\ref{fig:P_vs_m}(b)), 
suppressing the naively expected pairs formed by the majority components of the fermions. 

{\bf Conclusions:} In this work, we have investigated 
possible quintet-pair formation in the system of $F=3/2$ cold atoms in
one-dimensional optical traps at quarter filling via large-scale, high
precision DMRG simulations  of various correlation functions.  
We have found that 
sufficiently strong spin-imbalance
can stabilize different 
exotic quintet superfluid states, where pairs with different magnetic moments  coexist. 
We have found that for large magnetizations the dominant superfluid instability is 
determined by the most populated fermion components. For moderate 
magnetization, however, a different behavior was found: in the SF$_5(2,-2)$ phase 
the correlation function of pairs with 
the largest spin projections, $m=\pm 2$, show the slowest decay, which  probably indicates the emergence 
of an effective antiferromagnetic exchange between the pairs suppressing the quasi-long 
range order of all  other quintet pairs.

Quantum degeneracy of spin-3/2 fermionic atoms could probably be realized experimentally 
since several atoms, e.g., $^{132}$Cs, $^9$Be, $^{135}$Ba, $^{137}$Ba or
$^{201}$Hg (see Ref.~\onlinecite{lewenstein07a, wu10})  have $F=3/2$ as lowest hyperfine manifold. 
Higher spin fermion mixtures have already been realized very recently~\cite{taie10a}, 
and these higher spin systems might also show similar instability against quintet superfluidity as
found in this paper for spin-3/2 fermions. 
The interaction between alkaline earth atoms  or between atoms having similar orbital structure, i.e.,
a closed outer orbit, have SU(N) symmetry with very good accuracy.
The interaction between a series of spinor bosonic isotopes of alkali atoms 
also turned out to be nearly SU(N) symmetric.\cite{ho98a}
It is expected that the superfluid state found along the SU(4) line
($g_0=g_2$) is relatively easy to realize. Nevertheless,
the rapid development of experimental techniques with
ultracold atoms raises the hope that quantum degeneracy of  multicomponent fermionic atoms 
with non SU(N) symmetric ground state will also be achieved in the near future.
 
There are several possibilities to probe these many-body correlation
effects and to detect the coexisting quintet pairs with different magnetic moments. 
The pair gap can be studied by
radio-frequency spectroscopy \cite{regal03a} or momentum-resolved
Bragg-spectroscopy \cite{ernst10a}, although these measurements 
have the disadvantage that the pair gap is the same for quintet pairs
with different $m$, since the $s$-wave 
scattering length in the quintet channel does not depend on $m$.
Magnetic moment of the pairs can be   measured independently from 
the pair gap, e.g., via a Stern-Gerlach-like experiment by applying
inhomogeneous external magnetic field. 

{\bf Acknowledgments:} This research was supported in part by the Hungarian Research Fund (OTKA) under Grant No. K68340,  
K73455 and K100908. \"O. L. acknowledges support from the Alexander von Humboldt foundation. 
The authors acknowledge
computational support from Philipps Universit{\"a}t, Marburg and the Dynaflex Ltd.


\begin{thebibliography}{99}



\bibitem{lewenstein07a} M. Lewenstein {\it et al.},  Adv. Phys. {\bf 56}, 243 (2007).


\bibitem{wu2003a} C. Wu, J-P. Hu, and S-C. Zhang, Phys. Rev. Lett. {\bf 91}, 186402 (2003). 

\bibitem{harada03a}  
K. Harada, N. Kawashima, and M. Troyer,  Phys. Rev. Lett. {\bf 90}, 117203 (2003); 
C. Honerkamp and W. Hofstetter,  Phys. Rev. Lett. {\bf 92}, 170403 (2004); 
H.~H. Tu, G.~M. Zhang, and L. Yu,  Phys. Rev. B {\bf 74}, 174404 (2006);
M. Hermele, V. Gurarie, and A.~M. Rey,  Phys. Rev. Lett. {\bf 103}, 135301 (2009); 
E. Szirmai, M. Lewenstein,  EPL {\bf 93}, 66005 (2011). 

\bibitem{gorshkov10a} 
M. A. Cazalilla, A. F. Ho, and M. Ueda, New J. Phys. {\bf 11}, 103033 (2009);
A. V. Gorshkov {\it et al.}, Nature Physics {\bf 6}, 289 (2010). 


\bibitem{buchta07a}
K. Buchta, \"O. Legeza, E. Szirmai, and J. S\'olyom,
 Phys. Rev. B {\bf 75}, 155108 (2007); E. Szirmai, \"O. Legeza, and J. S\'olyom,
 Phys. Rev. B {\bf 77}, 045106 (2008);
S. R. Manmana  {\it et al.},  Phys. Rev. A {\bf 84}, 043601 (2011).

\bibitem{wu05a} C. Wu,  Phys. Rev. Lett. {\bf 95}, 266404 (2005); C. Wu, Mod. Phys. Lett. B {\bf 20} 1707 (2006);
S. Capponi  {\it et al.},  Phys. Rev. B {\bf 75}, 100503(R) (2007); G. Roux, S. Capponi, P. Lecheminant, P. Azaria,  Eur. Phys. J. {\bf 68}, 293 (2009). 

\bibitem{rapp07a} \'A. Rapp {\it et al.},  Phys. Rev. Lett. {\bf 98}, 160405 (2007);
X. W. Guan, M. T. Batchelor, C. Lee, and J. Y. Lee, EPL {\bf 86}, 50003 (2009).

\bibitem{ho99a} T.-L. Ho and S. Yip, Phys. Rev. Lett. {\bf 82}, 247 (1999);
S. Yip and T.-L. Ho, Phys. Rev. A {\bf 59}, 4653 (1999). 

\bibitem{wu10}  C. Wu,  J. Hu, and S-C. Zhang, Int. J. Mod. Phys. B {\bf 24}, 311 (2010).

\bibitem{wu2003b} C. Wu Physics 3, 92 (2010). 

\bibitem{ferro} S. S. Saxena {\it et al.},  Nature  {\bf 406}, 587 (2000);
C. Pfleiderer {\it et al.}, Nature  {\bf 412}, 58 (2001).

\bibitem{pnictides} C. de la Cruz {\it et al.}, Nature {\bf 453}, 899 (2008);
Z. Ren {\it et al.}, Phys. Rev. Lett. {\bf 102}, 137002 (2009);
J. Paglione and R. L. Greene    Nature Physics   {\bf 6}, 645  (2010).

\bibitem{fulde64a} P. Fulde and R. A. Ferrell, Phys. Rev. {\bf 135}, A550 (1964); A. I. Larkin and Yu. N. Ovchinnikov, Sov. Phys. JETP {\bf 20}, 762 (1965).


\bibitem{fflo} 
K. Yang, Phys. Rev. B {\bf 63}, 140511(R) (2001); 
T. Mizushima, K. Machida, and M. Ichioka, Phys. Rev. Lett. {\bf 94}, 060404 (2005); 
A. E. Feiguin and F. Heidrich-Meisner Phys. Rev. B {\bf 76}, 220508 (2007);
G. Orso, Phys. Rev. Lett. {\bf 98}, 070402 (2007); 
M. Rizzi {\it et al.}, Phys. Rev.  B {\bf 77}, 245105 (2008); 
A. L\"uscher, R. M. Noack, and A. M. L\"auchli,  Phys. Rev. A {\bf 78}, 013637 (2008); 
F. Heidrich-Meisner, G. Orso, and A. E. Feiguin Phys. Rev. A {\bf  81}, 053602 (2010). 

\bibitem{batrouni07} G.~G. Batrouni, M.~H. Huntley, V.~G. Rousseau, and R.~T. Scalettar, Phys. Rev. Lett. {\bf 100}, 116405 (2008).

\bibitem{ovachkin} Y. Liao {\it et al.}, Nature {\bf 467}, 567 (2010).

\bibitem{3body} 
J. H. Huckans, J. R. Williams, E. L. Hazlett, R. W. Stites, and K. M. O’Hara,
Phys Rev. Lett. 102, 165302 (2009);
T. B. Ottenstein, T. Lompe, M. Kohnen, A. N. Wenz, and S. Jochim,
Phys. Rev. Lett. 101, 203202 (2008).

\bibitem{3body-kantian} A. Kantian, M. Dalmonte, S. Diehl, W. Hofstetter, P. Zoller, and A. J. Daley, Phys. 
Rev. Lett. {\bf 103}, 240401 (2009). 



\bibitem{clebsch}
The explicit form of the pairing operators is
\vskip -0.3cm
\[P_{00,i}^{\dagger}=c^{\dagger}_{3/2,i}c^{\dagger}_{-3/2,i}-c^{\dagger}_{1/2,i}c^{\dagger}_{-1/2,i}\]
\vskip -0.3cm
\[P_{20,i}^{\dagger}=c^{\dagger}_{3/2,i}c^{\dagger}_{-3/2,i}+c^{\dagger}_{1/2,i}c^{\dagger}_{-1/2,i}\]
\[~~~~P^{\dagger}_{2,\pm1,i}=\pm\sqrt{2}c^{\dagger}_{\pm3/2,i}c^{\dagger}_{\mp1/2,i} ~~P^{\dagger}_{2,\pm2,i}=\pm\sqrt{2}c^{\dagger}_{\pm3/2,i}c^{\dagger}_{\pm1/2,i}\]


\bibitem{dmrg} S. R. White, Phys. Rev. Lett. {\bf 69}, 2863 (1992); Phys. Rev.
B {\bf 48}, 10345 (1993); U. Schollw\"ock, Rev. Mod. Phys. {\bf 77}, 259 (2005).

\bibitem{taie10a} S. Taie {\it et al.}, Phys. Rev. Lett. {\bf 105}, 190401 (2010); 
H. Hara, Y. Takasu, Y. Yamaoka, J. M. Doyle, and Y. Takahashi, Phys. Rev. Lett. {\bf 106}, 205304 (2011).


\bibitem{ho98a}  Tin-Lun Ho, Phys. Rev. Lett. {\bf 81}, 742 (1998).

\bibitem{regal03a} C. A. Regal, C. Ticknor, J. L. Bohn, and D. S. Jin,  Nature {\bf 424}, 47 (2003);
C. Chin {\it et al.}, Science {\bf 305}, 1128 (2004).


\bibitem{ernst10a} P. T. Ernst {\it et al.},  Nature Physics {\bf 6}, 56 (2010).


\end{thebibliography}
\end{document}